\documentclass[%
 aip,
rsi,
 amsmath,amssymb,
 reprint,%
reprint,
superscriptaddress,
pra]{revtex4-1}
\usepackage{float}
\usepackage{graphicx}% Include figure files
\usepackage{dcolumn}% Align table columns on decimal point
\usepackage{bm}% bold math
\usepackage{float}
\usepackage[utf8]{inputenc}
\usepackage[T1]{fontenc}
\usepackage{mathptmx}
\usepackage{etoolbox}
\usepackage[center]{caption}
\makeatletter
\def\@email#1#2{%
 \endgroup
 \patchcmd{\titleblock@produce}
  {\frontmatter@RRAPformat}
  {\frontmatter@RRAPformat{\produce@RRAP{*#1\href{mailto:#2}{#2}}}\frontmatter@RRAPformat}
  {}{}
}%
\makeatother
\begin{document}

\preprint{AIP/123-QED}

\title{Compact in-vacuum gamma-ray spectrometer for high-repetition rate PW-class laser-matter interaction}
% Force line breaks with \\

\author{G. Fauvel}
\email{Gaetan.Fauvel@eli-beams.eu}
\affiliation{ 
ELI Beamlines Facility|The Extreme Light Infrastructure ERIC
Za Radnicí 835, 252 41 Dolní Břežany, Czech Republic%\\This line break forced with \textbackslash\textbackslash
}%
\affiliation{ Czech Technical University in Prague,
 Faculty of Nuclear Sciences and Physical Engineering , 115 19, Prague, Czech Republic}

\author{K. Tangtartharakul}
\affiliation{Center for Energy Research, University of California San Diego, La Jolla, CA 92093, USA}

\author{A. Arefiev}
\affiliation{Center for Energy Research, University of California San Diego, La Jolla, CA 92093, USA}

\author{J. De Chant}
\affiliation{ Lawrence Berkeley National Laboratory, 1 Cyclotron Rd, Berkeley, CA 94720, USA}
\affiliation{Also at Michigan State University, East Lansing, MI 48824, USA}
\iffalse
\author{Eric Esarey}
\affiliation{ Lawrence Berkeley National Laboratory, 1 Cyclotron Rd, Berkeley, CA 94720, USA}
\author{Cameron Geddes}
\affiliation{ Lawrence Berkeley National Laboratory, 1 Cyclotron Rd, Berkeley, CA 94720, USA}
\fi
\author{S. Hakimi}

\affiliation{ Lawrence Berkeley National Laboratory, 1 Cyclotron Rd, Berkeley, CA 94720, USA}

\author{O. Klimo}
\affiliation{ 
ELI Beamlines Facility|The Extreme Light Infrastructure ERIC
Za Radnicí 835, 252 41 Dolní Břežany, Czech Republic%\\This line break forced with \textbackslash\textbackslash
}
\affiliation{ Czech Technical University in Prague,
 Faculty of Nuclear Sciences and Physical Engineering , 115 19, Prague, Czech Republic}

\author{M. Manuel}
\affiliation{General Atomics, San Diego, CA 92186, USA}

\author{A. McIlvenny}
\affiliation{ Lawrence Berkeley National Laboratory, 1 Cyclotron Rd, Berkeley, CA 94720, USA}
\author{K. Nakamura}
\affiliation{ Lawrence Berkeley National Laboratory, 1 Cyclotron Rd, Berkeley, CA 94720, USA}

\author{L. Obst-Huebl}
\affiliation{ Lawrence Berkeley National Laboratory, 1 Cyclotron Rd, Berkeley, CA 94720, USA}

\author{P. Rubovic}
\affiliation{ 
ELI Beamlines Facility|The Extreme Light Infrastructure ERIC
Za Radnicí 835, 252 41 Dolní Břežany, Czech Republic%\\This line break forced with \textbackslash\textbackslash
}%

\iffalse

\author{C. Schroeder}
\affiliation{ Lawrence Berkeley National Laboratory, 1 Cyclotron Rd, Berkeley, CA 94720, USA}
\author{J. van Tilborg}
\affiliation{ Lawrence Berkeley National Laboratory, 1 Cyclotron Rd, Berkeley, CA 94720, USA}

\fi

\author{S. Weber}

\affiliation{ 
ELI Beamlines Facility|The Extreme Light Infrastructure ERIC
Za Radnicí 835, 252 41 Dolní Břežany, Czech Republic%\\This line break forced with \textbackslash\textbackslash
}%

\author{F.P. Condamine}
\affiliation{ 
ELI Beamlines Facility|The Extreme Light Infrastructure ERIC
Za Radnicí 835, 252 41 Dolní Břežany, Czech Republic%\\This line break forced with \textbackslash\textbackslash
}%

\date{\today}% It is always \today, today,

\begin{abstract}
With the advent of high repetition rate laser facilities, novel diagnostic tools compatible with these advanced specifications are required. This paper presents the design of an active gamma-ray spectrometer intended for these high repetition rate experiments, with particular emphasis on functionality within a PW level laser-plasma interaction chamber's extreme conditions. The spectrometer uses stacked scintillators to accommodate a broad range of gamma-ray energies, demonstrating its adaptability for various experimental setups. Additionally, it has been engineered to maintain compactness, electromagnetic pulse resistance, and ISO-5 cleanliness requirements while ensuring high sensitivity. The spectrometer has been tested in real conditions inside the PW-class level interaction chamber at the BELLA center, LBNL. The paper also outlines the calibration process thanks to a $^{60}$Co radioactive source. 
\end{abstract}

\maketitle

\section{Introduction}

In recent years, advancements in laser technology across the globe have led to an increase, not only in power levels, but also in repetition rates. Presently, several facilities are capable of delivering PW class laser pulses at intervals of a few minutes to seconds\cite{borneis_design_2021,hakimi_lasersolid_2022,burdonov_characterization_2021,doria_overview_2020,volpe_generation_2019}. For instance, the L3 laser at ELI Beamlines in Czech Republic can generate 0.5 PW shots at a frequency of 3.3 Hz \cite{borneis_design_2021}. Similarly, the BELLA PW laser at the Lawrence Berkeley National Laboratory (LBNL) can deliver 1 PW shots at a repetition rate of 1 Hz \cite{nakamura_diagnostics_2017}. This surge in statistical data presents exciting opportunities for the plasma community, but it also necessitates corresponding advancements in diagnostic tools.

 It is crucial that detectors are compatible with the new energy ranges of produced particles \cite{ridgers_dense_2012,zhu_dense_2016,norreys_laser-driven_2009,jansen_leveraging_2018,higginson_near-100_2018,jirka_qed_2017,wang_power_2020,stark_enhanced_2016} at these new intensity levels. In this range, the gamma-ray photons produced can provide insights into topics like the production of gamma-ray bursts in astrophysics \cite{piran_physics_2005,meszaros_gamma-ray_2006}, radioactive waste management \cite{galy_bremsstrahlung_2007}, and the production of matter through photon-photon collision as theorized by Breit-Wheeler \cite{breit_collision_1934,krajewska_breit-wheeler_2012,blackburn_nonlinear_2018,sugimoto_positron_2023,he_achieving_2022,he_dominance_2021,he_single-laser_2021}. There is also significant industry interest in the reliable and well-characterized production of high energy photons, which can penetrate matter more efficiently and be used as non-destructive probes \cite{ben-ismail_compact_2011,wu_towards_2018}.

While several methods exist to produce high energy photons \cite{giulietti_intense_2008,galy_bremsstrahlung_2007,ledingham_laser-driven_2010,cole_experimental_2018,nerush_gamma-ray_2014,nerush_laser-driven_2015}, the Direct Laser Acceleration (DLA) method \cite{stark_enhanced_2016,jansen_leveraging_2018,wang_power_2020,gong_direct_2020} has promising laser-to-photon conversion in the range of several percent with a low divergence of gamma-ray emission. The described method involves accelerating electrons to relativistic speeds within a medium of near-critical density, defined as a threshold density above which the incident laser is predominantly reflected. This electron acceleration is obtained thanks to the laser photons collision with the target. This ponderomotive force pushes electrons along the laser axis generating a strong azimuthal magnetic field. This magnetic field subsequently alters the electrons' trajectories, bending them in the transverse direction. Such bending of the trajectories induces a synchrotron-like emission. This leads to a bright, collimated emission of high-energy gamma-rays. At PW levels, gamma-rays up to a few MeV (for 0.5 PW lasers pulses) to hundreds of MeV (for 10 PW lasers pulses) can be produced\cite{jansen_leveraging_2018,li_ultrafast_2017}.

In order to conduct these experiments, we need to design a diagnostic tool suitable for these high energies. Adapting current photon spectrometer designs to this high-energy spectrum presents some challenges. For example, the design of X-ray spectrometers based on crystal diffraction \cite{young_high-resolution_1998, bragg_reflection_1913, renner_challenges_2019, condamine_high-resolution_2019} is unthinkable for these wavelengths ($\sim$ 0.1 Å) as no crystal with sufficiently small interplanar distance exists. This necessitates the creation of a new detector type. Research in this new energy range has been ongoing for several years, leading to the development of various detectors \cite{hannasch_compact_2021, behrens_tld-based_2002, chen_bremsstrahlung_2008, albert_angular_2013, henderson_ultra-intense_2014, horst_tld-based_2015, rhee_spectral_2016}. These detectors usually utilize a pattern of alternating absorbers and Imaging Plates (IPs). By adjusting the thickness and material type of the absorbers, the energy spectrum can be unfolded. However, these stacking calorimeters have a significant limitation linked to the passivity of the detector. IPs must be analyzed between shots, a time-consuming process reducing the advantage of new high-repetition rate lasers without accumulating over several shots.

In response to this challenge, several research groups have developed active gamma-ray spectrometer designs \cite{stransky_development_2021,istokskaia_experimental_2021,behm_spectrometer_2018}. By operating as both absorbers and detectors simultaneously, scintillators uphold the principle of energy deposition variation while also ensuring compactness and sensitivity. These spectrometers are usually located outside the interaction chamber, with the stack of scintillators placed directly behind a Beryllium window or a thick aluminum flange. However, this additional obstacle can emit supplementary radiation or particles creating difficulties in retrieving the original spectrum. By positioning the spectrometer inside the chamber, we can strongly reduce this parasitic signal.

Besides, as interaction chambers continue to increase in size, centimeter-scale scintillators fail to cover the entire opening angle of the gamma rays, which is typically from 30° ($\sim$ 500 mrad) for low energy photons to few degrees ($\sim$150 mrad) for the highest energy photons, as predicted for gamma flash experiments\cite{stark_enhanced_2016} . Although increasing the size of the scintillators is an option, it is impractical to create very large windows due to pressure and costs constraints. Additionally, this reduces the spectrometer's adaptability to position as the angular position will be only defined by the position of flanges and/or windows around the interaction chamber. As a result, we decided to focus on designing a spectrometer that can withstand the extreme conditions of an experimental chamber during PW-class laser shots. The chamber often houses multiple components such as large optics which can significantly limit the available real space for additional devices like detectors. To address this, we introduce the design of a highly compact gamma-ray spectrometer in this paper. The device has been engineered with dimensions of $20 \times 20 \times 25 \, \text{cm}^3$
, making it particularly well-suited for high-repetition-rate experiments and adaptable to the rigorous environmental conditions characteristics of a laser-plasma interaction chamber (EMPs, cleanliness, vacuum pressure, etc.). 

This manuscript describes the design of the high repetition rate spectrometer suited for the extreme constraints of the PW-level interaction chamber in Section \ref{Section:Design}.
Section \ref{Section:Unfolding} elaborates on the methodology employed for unfolding the gamma-ray spectrum from the acquired spectrometer images. This process is crucial for translating the raw data into a meaningful energy spectrum.
The paper follows with a discussion of the calibration process, outlining the strategies employed to align the experimental setup with the theoretical models. Finally Section \ref{Section:BELLA} reviews the installation and use of the spectrometer inside at the BELLA center, a PW-class level interaction chamber, at LBNL (California, USA).

\section{Design of a gamma-ray spectrometer with stacking scintillators}
\label{Section:Design}
In this manuscript, gamma rays are classified as photons with energy > 100 keV. To accommodate these high energies, the fluorescence properties of various materials can be used. Gamma-ray photons deposit some of their energy within these materials, thereby exciting them. The materials subsequently relax, emitting fluorescence photons generally within the optical wavelength range. This allows the use of a camera to collect them. By arranging scintillators in a stack configuration in the gamma propagation axis, the first layers act as detectors for lower energy gamma rays and at the same time serve as a filter allowing for the detection of only (ideally) higher energy gamma-rays in the deeper scintillators .

\begin{figure}[H]
    \centering
    \includegraphics[width = 0.42\textwidth]{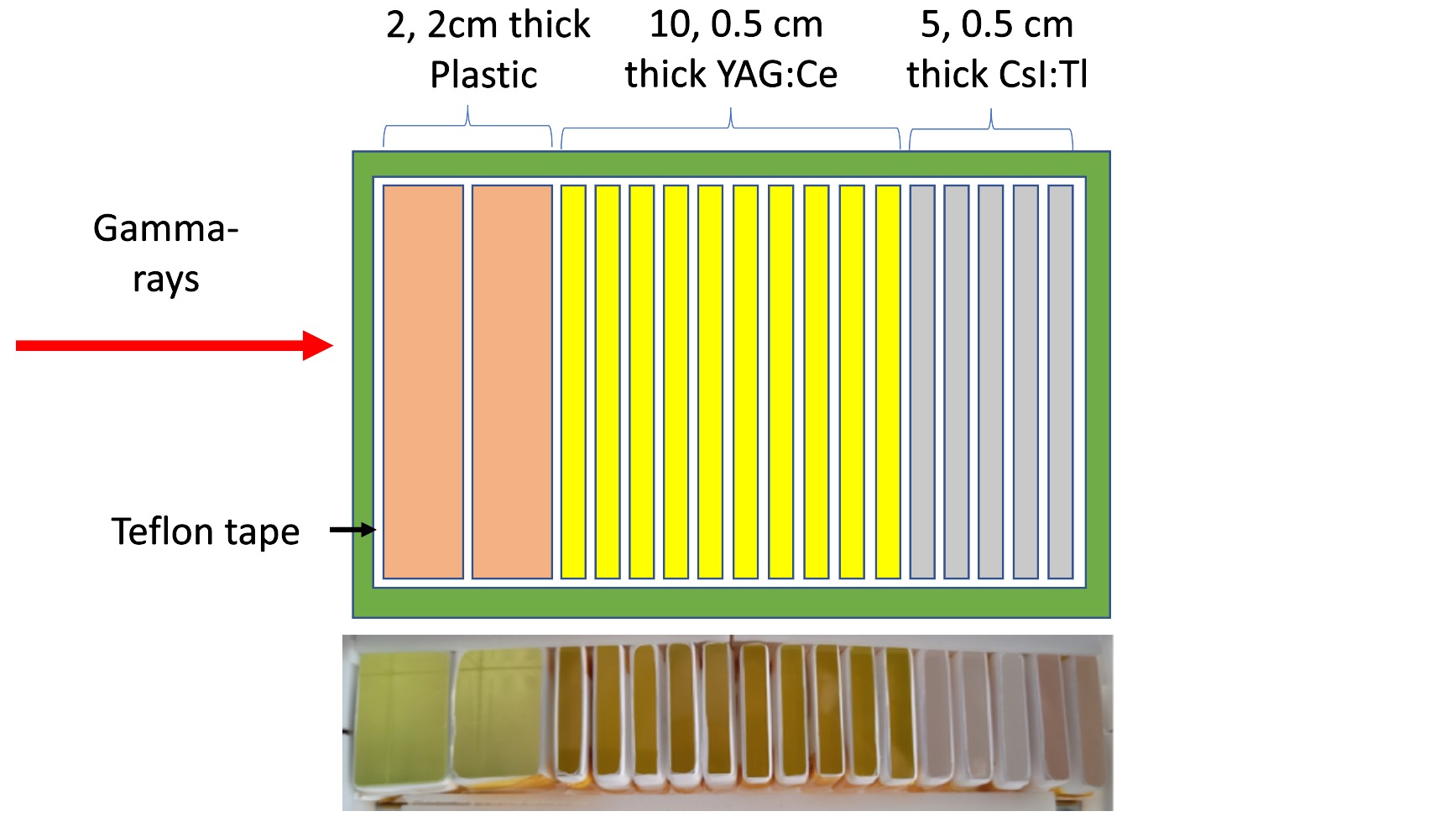}
    \caption{Scheme of the scintillator at the top and physical stack at the bottom. Each scintillator has the same transversal and vertical dimensions of 3 cm each, only the thickness varies. PTFE tape is wrapped around each scintillator on all faces except one facing towards the camera to increase photon collection, avoid the overlap of signals (optical photons going out through an adjacent scintillator) and allow some spatial separation for the camera aquisition.}
    \label{fig:Design_stack_0.5PW}
\end{figure}

Energy deposition is primarily dependent on three factors: the energy of the incident photon, the material of the scintillator, and its thickness. As we aim to determine the photon spectra, we have two adjustable parameters: the material and the thickness of the scintillator. Different types of scintillators offer different absorption coefficients for varying gamma-ray energies. The spectrometer designed in this study is intended to be used in different experimental setups covering various energy ranges. Consequently, we employ several types of scintillators to widen the energy range. The stack uses three different types of scintillators with various thicknesses related to their absorption coefficients. Plastic scintillators possess a low absorption coefficient, facilitating the detection of photons below 100 keV. YAG:Ce scintillators are more suitable for detection in the MeV range, and finally, CsI:Tl scintillators are ideal for energies above thanks to their short radiation length. The limited space within most interaction chambers led us to use a limited number of scintillators, in a layered configuration adapted to the expected gamma-ray spectrum. Here, it includes two 2 cm thick SP33 polystyrene (plastic) scintillators supplied by NUVIATech Instruments , ten 5 mm thick YAG:Ce scintillators, and five 5 mm thick CsI:Tl scintillators, both provided by Advatech UK Ltd. The arrangement of the scintillators is illustrated in Fig. \ref{fig:Design_stack_0.5PW}. Note that this modular design allows for the individual scintillators to be readily interchanged, offering flexibility to accommodate a range of energy spectra as necessitated by different experimental requirements. The design prioritizes compactness to facilitate integration with various experimental setups. Nevertheless, this compactness comes at the cost of the configuration versatility. To mitigate this, one could envisage a spectrometer with a larger number of scintillators.

A critical factor in the spectrometer's performance is the decay time of the scintillators, particularly the CsI:Tl, which has the longest decay time of 900 ns. This characteristic time can limit maximum repetition rate that can be used without signal overlap. Even if the scintillator hardware is far above the required characteristics, the camera becomes the limiting factor for the repetition rate achievable. As an example the camera used here from Allied Vision, model "Manta Camera 507-B", can only reach up to 23 fps but can be easily fixed with a high speed camera. With the current setup operating at a repetition rate of 10 Hz, the system is within the limits, ensuring clear temporal separation of detection events.

\begin{figure}[H]
    \centering
    \includegraphics[width = 0.4\textwidth]{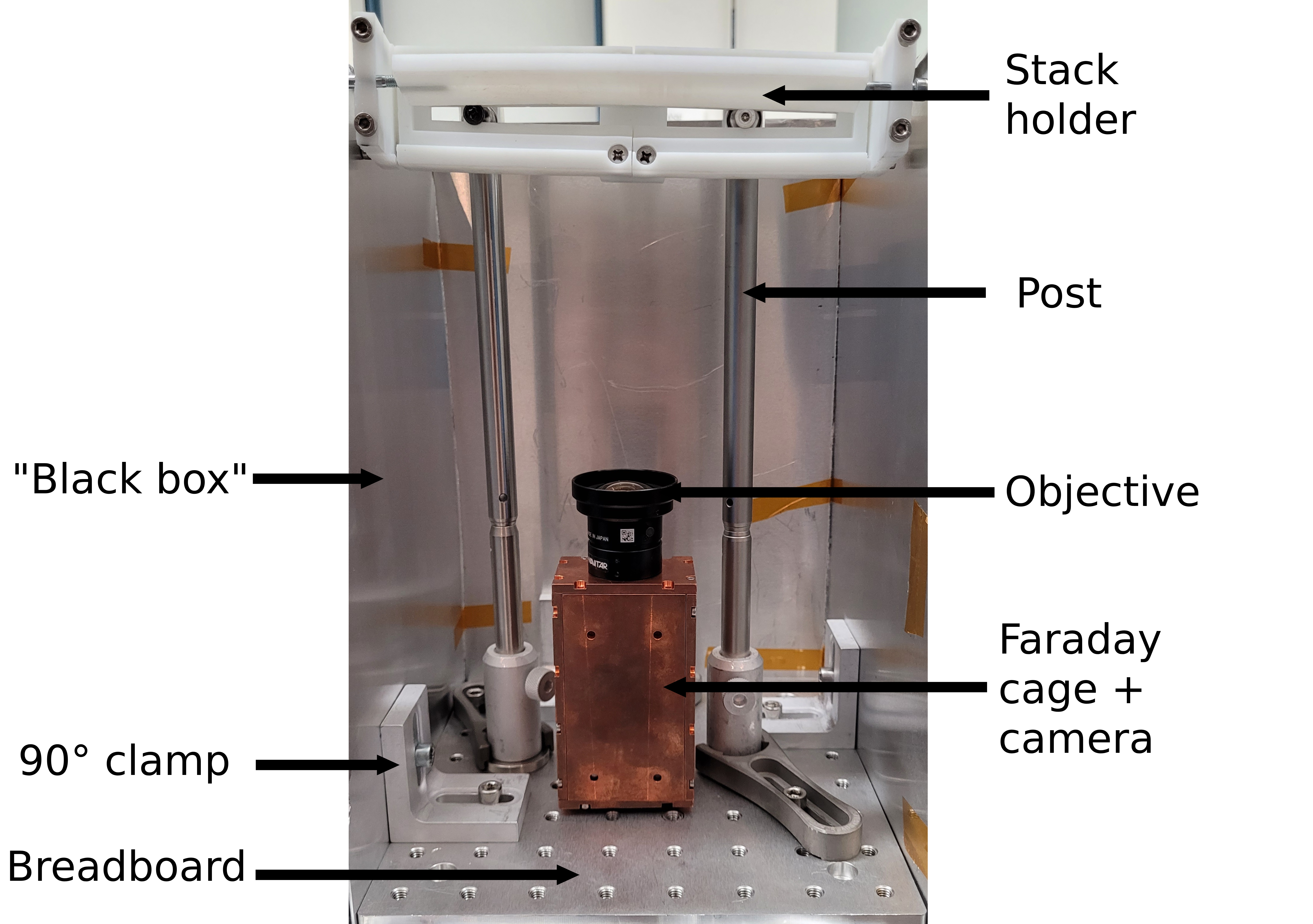}
    \caption{Spectrometer's assembly suited for installation inside a PW-class laser interaction chamber. One side and top panels have been removed for clarity. Stack configuration is presented in Fig. \ref{fig:Design_stack_0.5PW} }.
    \label{fig:Design_vacuum}
\end{figure}

The fluorescence photons from the scintillators are collected using an optical camera combined with an f/2.8 objective to have a large field of view with a short focal distance as the spectrometer height is only 20 cm. All faces are sanded except one polished on the smaller scintillators and all faces are polished on the larger ones. Each scintillator is wrapped in PTFE tape, leaving the polished face exposed, to enhance photon collection towards the camera while maintaining sufficient spatial spacing for scintillator differentiation. To shield the camera from Electromagnetic Pulses (EMPs) emitted during laser-plasma interaction, a 5 mm copper Faraday cage is installed around it. To protect the camera from the plasma self-emission (and any other parasitic light), a box composed of 2 mm aluminium sheets held by 90° clamps is positioned around the spectrometer. Using appropriate opto-mechanical elements, we meet cleanliness requirements, achieving a pressure of 10$^{-6}$ to 10$^{-7}$ mbar inside the interaction chamber and passing Residual Gas Analysis (RGA) tests. The spectrometer's assembly, suited for installation within the chamber, is presented in Fig.\ref{fig:Design_vacuum}.

Having underlined the hardware components of the detection system, attention must now turn to the process of data interpretation. The raw data captured by the camera requires a transformation to yield a meaningful gamma-ray spectrum. This transformation process, commonly referred to as "unfolding," is essential to convert the digitized image data into a physically interpretable format. Unfolding is a pivotal step in the experimental pipeline as the quality of the algorithm directly impacts the accuracy and resolution of the resultant gamma-ray spectrum, thereby influencing the interpretation of the underlying physical phenomena. Given the complexity of the data and the risk for various sources of noise and distortion, the algorithm applied must be robust and meticulously validated to ensure reliable results. The method is presented in the next section.

\section{Methodology for Unfolding}
\label{Section:Unfolding}

In order to accurately reconstruct the energy spectrum of the particle source under investigation, several computational and data analysis techniques must be employed on the data captured by the spectrometer. This data is acquired using a camera resulting in an image of the scintillators integrated over the interaction time. This image serves as a raw source of information and is directly correlated to the energy deposited within various scintillation materials. The latter can be calculated thanks to a Monte-Carlo code, taking as input the spectrometer's configuration and particle's energy spectrum.  It should be noted that the pixel value is a function of not only the particles' inherent energy and type but is also modulated by a set of calibration factors.

These calibration factors include a variety of conversion factors, each contributing to the overall signal transformation process from the Monter-Carlo simulation output to a camera pixel value equivalent. These factors are the intrinsic light yield of the scintillation materials utilized directly dependant on the material but can also vary depending on the manufacturing of the scintillator crystal. The angular dependencies affect photon collection efficiencies, for example, scintillators on the left of the field of view will have fewer photons collected by the objective than the scintillators in the center (for a same initial emission). The quantum efficiency of the camera, which is responsible for converting the collected photons into a digitized pixel array, can have different sensitivities for different wavelengths. This overall factor allows a simple linear relationship between the Monte-Carlo simulation output and the camera value. This conversion factor, unique for each scintillator, is defined through the calibration process, detailed in Section \ref{Section:Calibration}, thereby enabling the conversion of the scintillation image to meaningful particle energy. Equation \ref{eq:RLY} is a simple equation for these calibration factors.

\begin{equation}
\begin{array}{c c}
     \text{Camera pixel value} = & \text{Monte-Carlo output }\\
     &\times \sum \text{Calibration factors} 
\end{array}
\label{eq:RLY}
\end{equation}

To calculate the energy deposition in the scintillators, we employ a Monte Carlo simulation approach using the FLUKA code \cite{ahdida_new_2022,battistoni_overview_2015}. This is complemented by the FLAIR software \cite{vlachoudis_flair_2009}, a visual interface used to effectively configure and visualize the FLUKA simulations. Together, they provide a robust framework for rebuilding the experimental setup in a virtual environment, thereby allowing us to account for material-specific interactions, scattering, and other complex phenomena that influence the observed data.

Once the simulation environment is established, the energy spectrum of the particle source is iteratively adjusted. The aim is to minimize the divergence between the simulated and experimental data, subject to the modulation imposed by the calibration factor of the scintillators used. It is critical to note that while this is a widely used approach, alternative methodologies exist that might offer advantages in specific applications. For instance, machine learning techniques like neural networks or statistical methods such as Bayesian unfolding algorithms can also be employed to unfold the energy spectrum. Although these advanced techniques are beyond the scope of this work, they represent important directions for future research.

\section{Calibration using a $^{60}$Co radioactive source}
\label{Section:Calibration}
One of the most critical requirements for calibration is the selection of an appropriate radiation source. The ideal source should possess a well-defined and accurate energy spectrum, with photon energies exceeding the 500 keV range and a sufficiently high photon flux to ensure statistical reliability.

In our specific case, we opted to use sources with activities in the TBq range. After careful consideration, a $^{60}\text{Co}$ source was selected for its suitability for our experiment. This source has an activity of 0.185 TBq and emits photons with two dominant photopeaks at energies of 1.173 MeV and 1.332 MeV. The experimental setup was placed at a distance of 1 m from the source which has an opening angle of 11° ($\sim$ 200 mrad).

\begin{figure}[H]
    \centering
    \includegraphics[width = 0.425\textwidth]{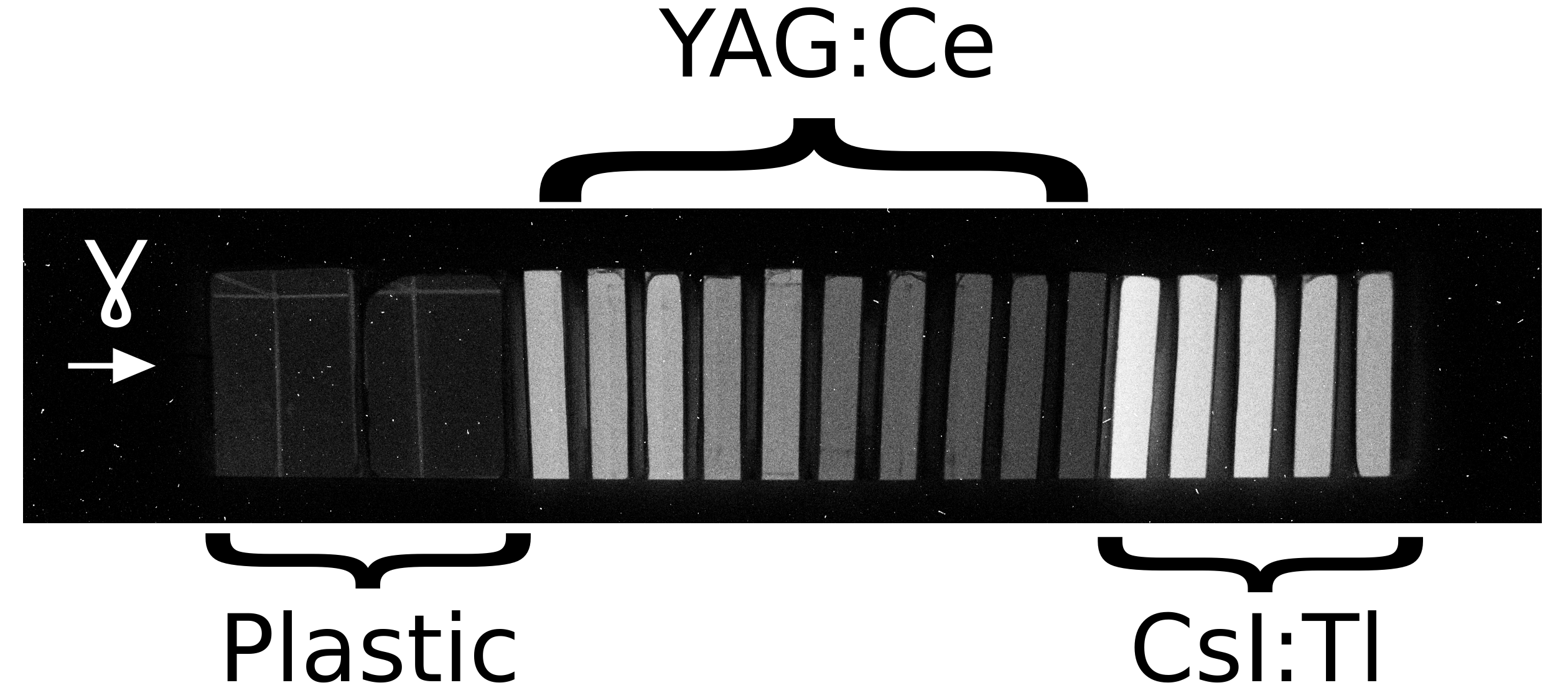}
    \caption{Observed image using a 60-Co calibration source at 1m of the detector. The camera was set to a gain of +40 db and 0.5 s exposure time.. Three different types of scintillators are used, explaining the abrupt changes in scintillation along the propagation axis. The radiation comes from the left with the source at 1m with an opening angle of 11°.The contrast of the camera image was enhanced.}
    \label{fig:Calibration}
\end{figure}

Data acquisition was executed under varying conditions of gain and exposure time to optimize the sensitivity and dynamic range of the camera. A typical scintillation image obtained from this setup is shown in Fig. \ref{fig:Calibration}.

Upon the successful completion of the data acquisition phase, we proceeded to start the comparative analysis between our experimentally derived data and the simulations generated from FLUKA. This meticulous comparison was critical for the extraction of the calibration factors key metrics defined in Equation \ref{eq:RLY}.  

\begin{figure}[H]
    \centering
    \includegraphics[width = 0.45\textwidth]{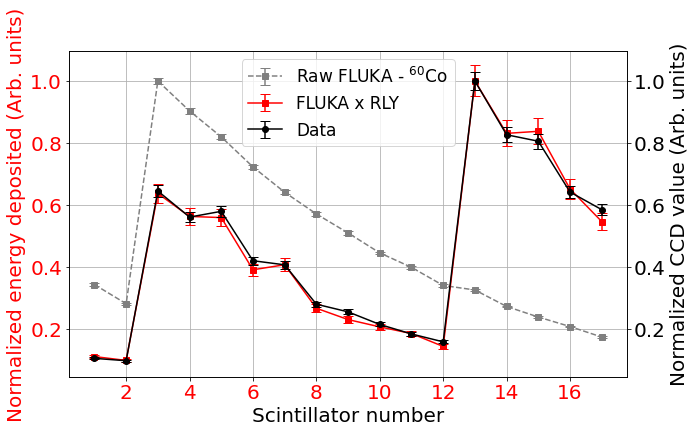}
    \caption{Comparison between camera output image (black full line) and FLUKA simulation for a 0.185 TBq $^{60}$Co calibration source (gray dashed line) used for calibration factors calculation. In red is the FLUKA output times the calibration factors, this curve only takes into account the photon spectra. All data is normalized to 1.}
    \label{fig:Calibration_RLY}
\end{figure}

To further refine the data and mitigate inconsistencies or inhomogeneities, we applied a statistical treatment to the acquired pixel values. Specifically, the mean pixel value was calculated for each individual scintillator in the stacked array reducing the number of outliers which can make comparison with simulations more difficult. The processed mean pixel values are plotted against the FLUKA simulation results, as illustrated in Fig. \ref{fig:Calibration_RLY}.

These quantifications, which represent the photon to pixel value conversion efficiency, are crucial for the accurate interpretation of the energy spectrum deduced from the observed data. This means that every FLUKA output will be multiplied by the calibration factor specific to each scintillator before trying to compare it to the experimental data. This way it is possible to retrieve the original data using the FLUKA output and calibration factors as presented in Fig. \ref{fig:Calibration_RLY}. As the calibration source produces some electrons, they generate low energy photons by Bremsstrahlung in air. In order to observe the impact of these electrons, the red curve shows the energy deposition pattern when only the primary photons of the radioactive source are taken into consideration.

\section{Preliminary Results from Petawatt-Class Laser Interactions in the BELLA PW iP2 chamber at LBNL}
\label{Section:BELLA}

In order to test the spectrometer in real conditions, we participated in an experiment at the BELLA Center at LBNL. Using a PW laser, shots on foam target, were performed for radiation regeneration \cite{hakimi_laser-ion_2024}. The spectrometer was successfully implemented and the corresponding acquired data is discussed hereafter.

\begin{figure}[H]
\centering
\includegraphics[width = 0.35\textwidth]{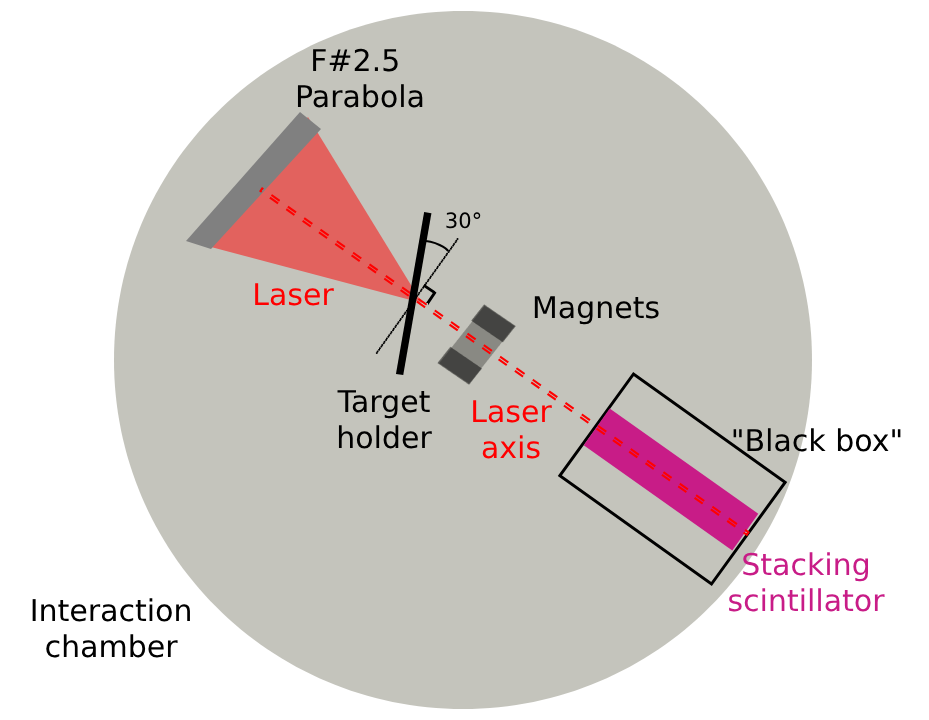}
\caption{Schematic representation of the experimental setup in the iP2 chamber of the BELLA PW laser. The gamma-ray spectrometer is situated along the laser axis at $\sim$50cm from the interaction point, following magnets intended to divert electrons away from the gamma-ray spectrometer. Target is shot at a 30 degree incidence angle. Dimensions are not to scale.}
\label{fig:BELLA_experiment}
\end{figure}

\begin{figure}[H]
\centering
\includegraphics[width = 0.35\textwidth]{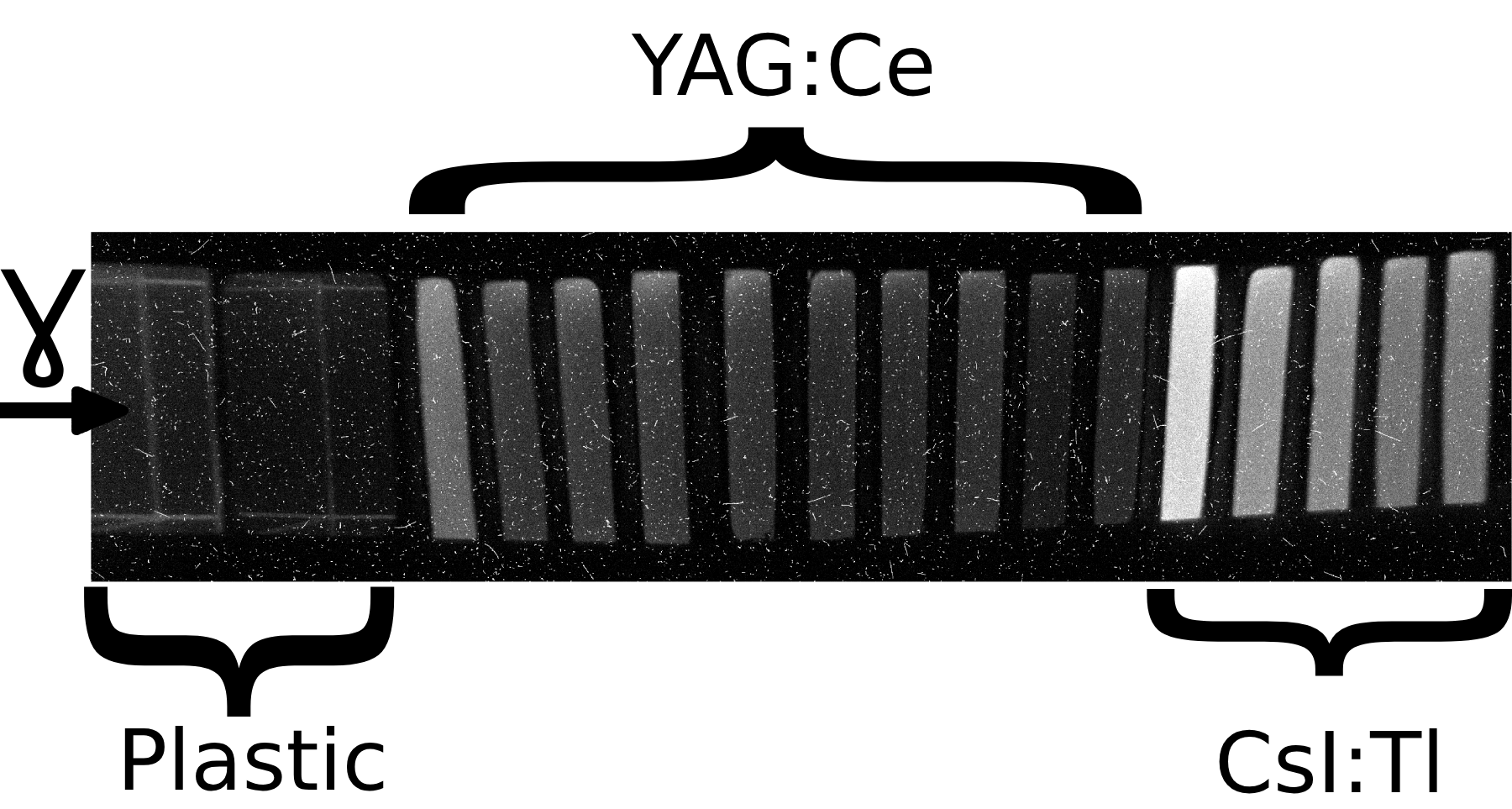}
\caption{Camera image captured during a PW-class level interaction with a foam target. The left black arrow represents the gamma-rays direction of propagation.}
\label{fig:BELLA_data}
\end{figure}

The experimental configuration is illustrated in Fig.\ref{fig:BELLA_experiment}. Utilizing a $\sim$ 50 fs pulse, which was focused by an F/2.5 OAP\cite{obst-huebl_high_2023,hakimi_lasersolid_2022}, we achieved a focal spot size of $\sim$ 3$\mu$m at $\frac{1}{e^2}$and delivered an energy of $\sim$ 4 J onto foam targets of density $\sim$ 20 mg/cc and $\sim$ 40$\mu$m thickness. This resulted in the emission of photons, which were directed along the laser axis where the spectrometer was positioned. The experiment yielded multiple images, an example is shown in Fig. \ref{fig:BELLA_data}. These images show that acquiring gamma-ray data inside a PW-class interaction chamber with active elements is possible despite EMPs and cleanliness requirements.

\section{Discussion}

To conclude, in this manuscript an active gamma-ray spectrometer suited for use inside a PW-class laser interaction chamber is presented for gamma-rays detection from 100 keV to 10 MeV. This spectrometer offers several advantages in its design. First by using a stack of different scintillators it can reach high repetition rate of up to 10 Hz which can be easily increased by upgrading the camera model to a faster one. This design is modular meaning each scintillator can be easily replaced for maintenance or swapped to meet the needs of energy analysis or precision.

The spectrometer's compatibility with the extreme conditions of a PW-class laser interaction chamber, such as electromagnetic pulse (EMP) resistance, vacuum pressure tolerance, and maintenance of cleanliness, is ensured through a careful selection of materials and design.

This spectrometer has undergone a calibration process utilizing a radioactive source of $^{60}$Co. We used the observed data compared to simulations to calculate the calibration factors used as conversion factors between simulations to real-world data. Even if the configurations can be easily adapted to the experimental needs, due to an unavoidable variation in the photometric response, surface sanding, and efficacy of wrapping, when scintillators are interchanged or move locations within the diagnostic, new calibration factors are needed to accurately interpret data from laser experiments. The simulations performed thanks to the Monte-Carlo code FLUKA, allows to calculate the energy deposition of any spectrum shape. 

The spectrometer has been successfully deployed during an experimental campaign at the BELLA Center of LBNL. The obtained results demonstrate the feasibility of recording gamma-ray data inside a PW-class interaction chamber with active elements which had not been, to the best of our knowledge, done yet.
Then, by benefiting from high-repetition rate operation at state-of-the-art facilities the errors and uncertainties of the measurement will be iterated down by recording more data sets for different experimental conditions and stack configurations. This possibility, impossible with low-repetition rate lasers and diagnostics, will allow to compare 3D PIC simulations to the experimental data and lead to a better understanding of the physics that is at play during DLA and other gamma-ray generation processes.
\begin{acknowledgments}

We would like to thank V. Istokskaia , L. Giuffrida and B. Lefebvre, R. Versaci,  C. Lacoste, for fruitful conversation regarding the design and unfolding method. We would like to also thank 
The National Radiation Protection Institute (SÚRO) in Prague, Cz for their help in the calibration process.

We wish to acknowledge the support of the National Sci-
ence Foundation (NSF Grant No. PHY-2206777) and the Czech Science Foundation
(GA ČR) for funding on project number No. 22-42890L in the frame of the National Science Foundation–Czech Science Foundation partnership.

The work conducted at BELLA was supported by the U.S. DOE Office of Science Offices of HEP and FES (incl. LaserNetUS) under Contract No. DE-AC02-05CH11231, the Defense Advanced Research Projects
Agency via Northrop Grumman Corporation, and the U.S. DOE FES Postdoctoral Research Program, administered by ORISE under contract DE-SC0014664.
\end{acknowledgments}

\section*{Data Availability Statement}

The data that support the findings of this study are available from the corresponding author upon reasonable request.

\section*{References}

\bibliography{RSI}% Produces the bibliography via BibTeX.

\end{document}